\documentclass[
    ,final            % use final for the camera ready runs
  ]
  {aipproc}

\layoutstyle{6x9}

\def\Msol{M_{\odot}}

\def\keV{\, {\rm keV}}
\def\apj{ApJ}
\def\mnras{MNRAS}
\def\aap{A\&A}
\def\araa{Ann.\ Rev.\ A\&A}

\begin{document}

\title{The Monster's Fiery Breath and its Impact on Galaxy Formation}

\classification{98.65.Hb; 98.62.Ai}
\keywords      {Galaxies: Formation; Galaxies: Clusters}

\author{Richard Bower}{
  address={University of Durham, South Road, Durham, UK}
}

% \author{<author2>}{
%   address={<common address for author2 and author3>}
% }
% 
% \author{<author3>}{
%   address={<common address for author2 and author3>}
%   ,altaddress={<author1 address>} % additional visiting address
% }

\begin{abstract}
My aim in this talk is to make clear that there are two sides to
galaxy formation: the properties of the galaxies themselves, and the 
properties of the material that is left over from the galaxy formation
process. To date, galaxy formation studies have focused on correctly
predicting the properties of galaxies, and I will review the tremendous
level of success in this area. However, these models usually ignore the 
``flip side'' of galaxy formation: the intergalactic medium and the 
intra-group/intra-cluster medium (ICM).  
Yet, Chandra and XMM have given us a good
view of the ICM and their results present an equally important
challenge for theoretical models. I will show that this challenge is far 
from easy to meet, but describe the Bower et al 2008 model of galaxy formation
which successfully combines both sides of the observational constraints.
\end{abstract}

\maketitle

%%%%%%%%%%%%%%%%%%%%%%%%%%%%%%%%%%%%%%%%%%%%
%% MAINMATTER
%%%%%%%%%%%%%%%%%%%%%%%%%%%%%%%%%%%%%%%%%%%%

\section{Galaxy Formation}

Until recently, there were three longstanding problems that galaxy formation 
models struggled to explain. (1) The puzzling inefficiency of galaxy 
formation: only $\sim 10\%$ of the baryonic matter has cooled and been locked into 
stars and cold gas within galaxies (Balogh et al.\ 2001). (2) The sharp break at the bright end 
of the luminosity function: the abundance of galaxies drops rapidly at high mass,
much more rapidly than can be explained by the increase in cooling
time with system mass (White \& Rees 1978, Benson et al.\ 2003). (3) observational evidence 
that galaxy formation is ``anti-hierarchical'': if the formation and assembly of
galaxies mirrored the formation of cold dark matter haloes, we would expect the largest
galaxies to have formed most recently; the observational data points to the 
opposite conclusion (eg., Glazebrook et al.\ 2004). Now, a surprising solution to these problems has emerged:
galaxy formation appears to be intimately coupled to the growth of black holes, with
AGN playing a key role in preventing cooling in the most massive galaxy haloes 
(eg.\ Croton et al.\ 2006; Bower et al 2006, hereafter B06).

The idea has been inspired by observational evidence of a ``cooling flow crisis''. Although the 
cooling times at the centers of clusters are short, X-ray spectroscopy has shown that the
little cool gas (with temperature below a third of the virial temperature) is 
present (eg.\ Peterson et al.\ 2003). 
This suggests the presence of a heat source that offsets to radiative losses of the system.
Moreover, a suitable heat source (at least in terms of its energetics) is usually seen to 
be present in the form of a central radio galaxy and/or cavities in the X-ray
surface brightness (presumably filled with tenuous relativistic plasma) (Binney \& Tabor 1995;
Churazov et al.\ 2001). 
Estimates of the injected energy seen in the cavities appear to match the thermal energy 
loss remarkably
well (eg.\ Dunn \& Fabian, 2008), although coupling the energy source to the cooling plasma in 
an optimal way remains a complex issue (cf.\ McNamara \& Nulsen 2007, and this meeting).

\begin{figure}
\includegraphics[width=8.6cm, angle=270]{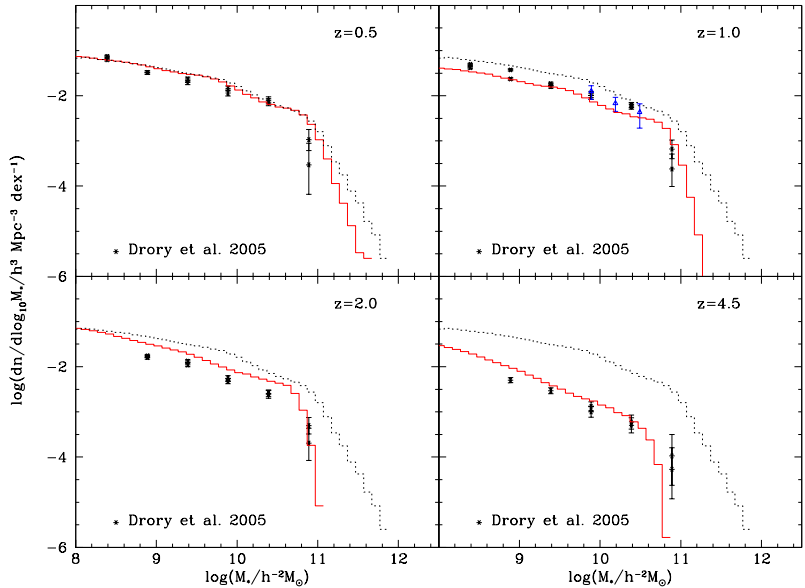}
\caption
{The evolution of the stellar mass function in the Bower et al.\ 2006
galaxy formation model. The model predictions
are shown by the red solid lines at the redshifts indicated in the
legend. The dotted line in each panel corresponds to the model stellar
mass function at $z=0$. The observational estimates from Drory et
al. (2005), based on photometric redshifts, are shown by symbols with
error bars (similar results are obtained
from spectroscopic surveys). The agreement between the observed
evolution and that predicted by the model is striking.}
\label{fig:b06_fig5}
\end{figure}

Although ``the monster's fiery breath'' was originally proposed as a solution to the cooling 
flow crisis on cluster scales, it turns out that this provides exactly the mechanism that is
needed to solve the three problems of galaxy formation. While different authors have implemented
the effect of AGN in very different ways, the effect is similar. By suppressing the cooling of 
gas in haloes of mass $10^{12}\Msol$ (interestingly comparable to the mass of the Milky Way's
own halo),  the model achieves a sharp roll over of the the galaxy luminosity function.
By construction, the ``radio mode'' of AGN feedback is only effective in 
hydrostatic haloes (eg.\ Birnboim \& Dekel 2003). 
Combined with effective supernova feedback in lower mass galaxies, this suppression also leads 
to a much lower global fraction of stars and to a good match to the shape of the cosmic star 
formation history.  Moreover, ``breaking'' the link between halo growth and the 
formation of galaxies naturally leads to an ``anti-hierarchical'' sequence of 
galaxy formation.  Since, the most massive haloes gain mass quickly,
star formation is truncated by the AGN at an earlier epoch (higher redshift) than in lower
mass haloes. As a result, the most massive galaxies have the 
oldest stars, just as the observational evidence shows (Fig.~\ref{fig:b06_fig5}).

\section{The flip-side of galaxy formation}

The success of galaxy formation models is usually measured by 
their ability to match key observational properties of the galaxy 
distribution. However, in successful galaxy formation models, the vast majority 
of baryons remain in diffuse form, either because they 
are unable to condense out of the intergalactic medium (for example, because
their host dark matter halos are too small to resist heating from the diffuse
inter-galactic background radiation) 
or because they are ejected from the star forming regions
of galaxies by strong feedback (White \& Frenk 1991). 
In general, it is difficult to observe 
the intergalactic medium directly; however, within groups 
and clusters of galaxies the intergalactic medium becomes sufficiently hot (and dense)
that it can be observed at X-ray wavelengths.

There is a long history of work attempting to explain the observed properties 
of the ICM of groups and clusters 
(e.g., Kaiser 1991; Evrard \& Henry 1991; Tozzi \& Norman 2001; McCarthy et al.\ 2008). 
Generally, it has been concluded that the observed properties cannot be
explained by the gravitational collapse of dark matter haloes alone: the energetics 
of observed clusters and the scaling of the X-ray emission with system temperature
suggest that an additional heat source is required. This
is most efficiently achieved by heating the ICM prior to its collapse so that 
a high minimum adiabat is set, resisting the gravitational compression of the system.
Such preheating models have been explored extensively in the literature.
However, the source of the 
preheating energy is rarely explicitly modeled and the scaling of system entropy with
mass ($K_{\rm vir}\propto T_{\rm vir}\rho_{\rm vir}^{-2/3}\,$) 
makes it difficult to simultaneously preheat the IGM to a sufficiently high 
adiabat that it is able explain the properties of galaxy clusters and yet retain sufficient
low entropy gas in lower mass haloes to obtain a realistic galaxy population. This is a 
generic problem --- few models attempt to explain the properties 
of the ICM while simultaneously accounting for the observed properties of galaxies. 

In view of the above difficulties, it is useful to take a step back from the problem.
If the properties of galaxies and the low observed stellar mass fraction are set aside,
cooling provides an appealing explanation for the observed scalings of the ICM
(Voit \& Bryan 2001). Because the cooling time is closely
related to the adiabat (or entropy) of the gas, lower mass systems (with lower
characteristic entropy) tend to cool out a larger fraction of their ICM. This is 
sufficient to reproduce many of the observed trends in X-ray properties, but the 
implied stellar fractions are much larger than those observed.
This suggests that a simpler alternative to the pre-heating model is worth further
investigation: we need to arrange for feedback to eject much of the cooling gas from the 
system before allowing it to form stars. We propagate the ejected gas fractions through the 
merger hierarchy so that the scheme has elements in common with the preheating
scenario discussed above. However, since gas is ejected in 
virialised haloes by `in-situ' heating, it has none of the energetic efficiency of the 
pre-heating scenario and the required energy injection will inevitably be large.

As we have discussed, the over-cooling problem can be 
resolved by including a ``radio mode'' of AGN feedback in the models
(in addition to the supernova-driven winds).
However, in order to explain the properties of the ICM, we need to take the process a step further.
We consider the possibility that 
sufficiently massive black holes not only prevent cooling in their host haloes, but
may also inject sufficient energy to expel gas from the halo. As gas is expelled, the 
central density drops, the cooling time becomes longer and the cooling rate, 
and hence energy feedback, become smaller. The system will move to a new lower density 
configuration where the energy feedback just balances the cooling rate. The concept
is appealing since the final configuration is set by the cooling time in the 
halo, while the ejection of gas avoids the excess production of stars. It combines
the simplicity of the scheme suggested by Voit \& Bryan (2001) while offering the
potential to give a good match to observed galaxy properties. The model allows us
to propagate the effects of heating at early epochs to later times
(Wu et al.\ 2000), but it does not
implement ``preheating'' in the way envisaged by many previous papers. 
Nevertheless, an order of magnitude calculation shows that the model
is worth further consideration: the energy released in the formation of 
the black holes in a cluster of $5\times10^{14}\Msol$ is comparable to
the binding energy of the ICM.  At lower masses the black hole energy
is potentially able to expel a large fraction of the baryonic mass from the system.

In the following model, I take full account of the different channels for black hole mass 
growth (we assume that black hole growth occurs through the ``QSO mode'' does not provide
heat the ICM efficiently) and for the effect of heating in subhaloes that are subsequently 
accreted by the growing cluster.
The methods we adopt are semi-analytic and based on the techniques
described in detail in Cole et al.\ (2000) and Baugh (2006), and the results are in accord
with fully numerical simulations (McCarthy et al., in prep).

\section{The Problem with Galaxy Formation Models}

The relation between bolometric X-ray luminosity and system temperature is a basic
observational correlation that can readily be measured over a wide range of system
mass. We first compare the data with the model in the absence of any heating. 
We use the large data sample compiled by Horner (2001) as the basis of our 
comparison. The data spans a wide range of system temperatures and is relatively 
unbiased with respect to system surface brightness and
supplement this with group data from the GEMS project (Osmond \& Ponman 2004). 
For the model comparison, we compute the bolometric
X-ray luminosity and X-ray spectral temperature ($T_{\rm spec}$). 

Even though the model is highly successful at describing galaxy properties,
it fails to describe the ICM at all well.  The left hand panel of
Fig.~\ref{fig:LT} shows the L--T plot for the model in the absence of
AGN heat input.   Black and red solid lines show the
median relations for the model and data respectively. This discrepancy is expected
--- it is well known that the scaling expected from gravitational collapse
is unable to explain the observed L--T relation (e.g., Kaiser 1991; Henry \& Evrard 1991). 
An improved match requires that
we add energy to the lower mass systems so that their central densities are lower
and the model relation becomes steeper.

\section{A Combined Model of Galaxy Formation and the ICM}

\begin{figure}
  \includegraphics[width=6cm]{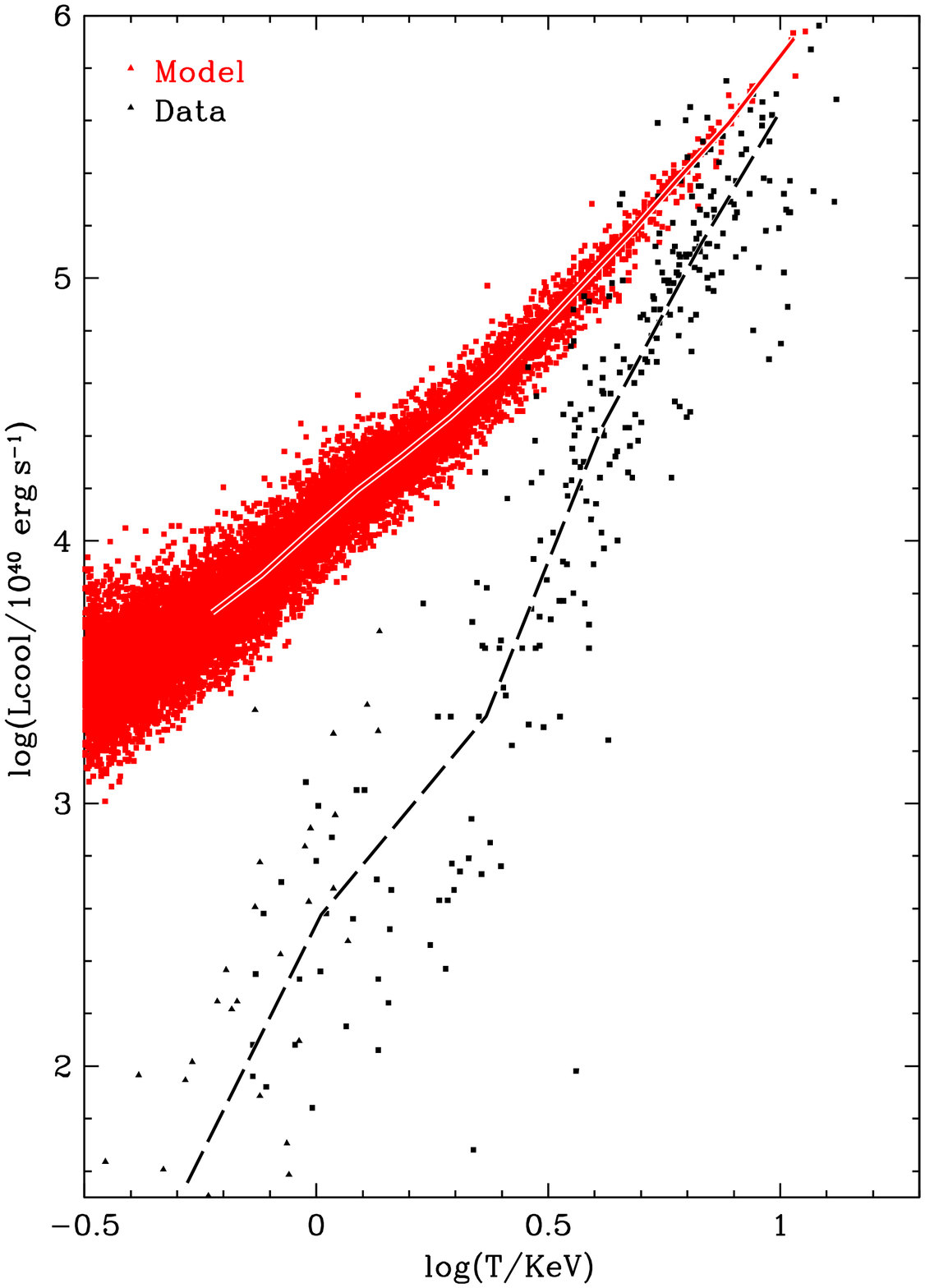}
  \includegraphics[width=6cm]{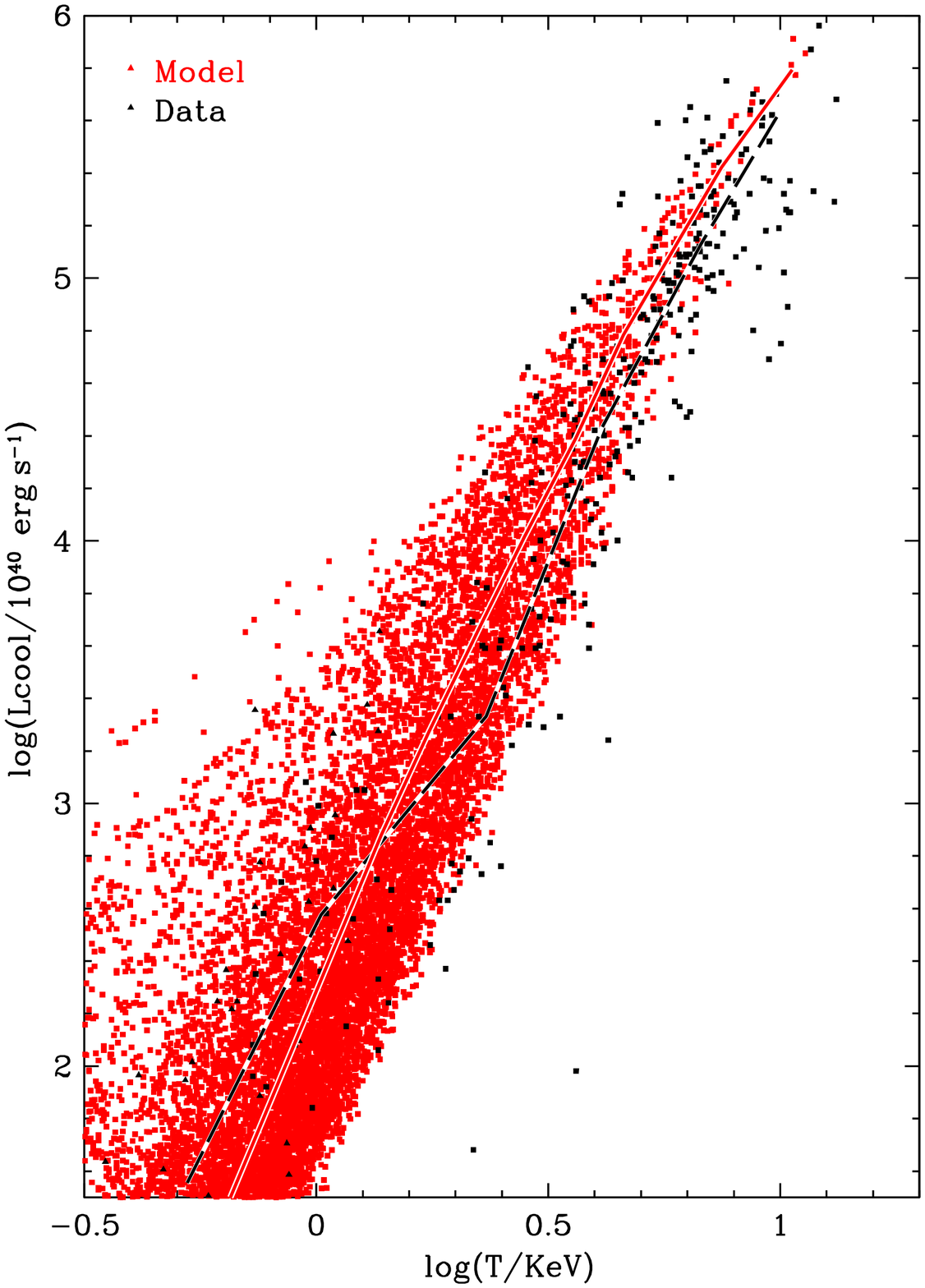}
\caption{A comparison of the bolometric Luminosity--Temperature relation for 
the Bower et al. 2006 (left panel) and Bower et al 2008 (right panel) models. 
Black squares present the observational data, while red points show the 
predicted X-ray luminosities of model haloes.
Red and black lines show the median relations for models and data respectively.
{\it Left Panel:} The original Bower et al.\ 2006: as expected from 
gravitational  scaling, the model relation is too shallow compared to 
the data. This illustrates the behaviour of models in which the AGN 
simply prevents gas cooling. {\it Right Panel}:
the revised model in which the AGN ejects gas from the system as well 
as preventing cooling. This results in a much improved match to the observational
data
} 
\label{fig:LT}
\end{figure}

We now consider the effect of including AGN heating in the model.
The effect of the gas ejection is to reduce the predicted luminosities of
lower mass systems. This occurs because the cooling time of the lower mass 
systems is shorter and thus they initially supply more material to the 
the AGN resulting in larger feedback energy (per unit gas mass). 
Hot gas is then ejected from the X-ray emitting region 
until the cooling rate (and the X-ray luminosity) of the system drops below a critical value. 
In contrast, the most massive systems have such long cooling times that little 
material is able to cool and they therefore retain close to the universal 
baryon fraction.

The L--T relation of the revised model is shown in the right-hand panel 
of Fig.~\ref{fig:LT}. The model
relation is now significantly steeper and in much better agreement with the observational
data. The model and observed median relations have similar slope and 
normalisation. In addition
to matching the median slope of the observed data, the model also shows a similar 
variation in the scatter along the relation. In particular, below a temperature of $\sim 3$ keV, 
the model points fan out to fill a triangular region of the L--T plot. The success of the 
model is evident in comparison with the left panel. 

The diversity of model groups in the low temperature region 
arises from the range of merger histories with data points lying towards
the high luminosity edge having recently undergone rapid mass growth. During the rapid growth
phase, gas mass is added to the system and the importance of the past heat input 
decreases relative to the gravitational potential of the new halo. Over time,
the AGN injects further energy in order to re-establish an equilibrium state
and the system will move towards the main relation.
Groups on the low luminosity side correspond to systems with unusually slow mass 
growth rates. This is a key success of the model --- reproducing the diversity of the 
X-ray properties of groups is something that must be added ``by hand'' to
preheating models. 

Of course, the heating model has a back-reaction on galaxy properties. 
The impact of the heating of hydrostatic haloes is significant, but
the other parameters of the model can be adjusted to ensure that a good match
to the luminosity function is still obtained. The model also
reproduces the observed black hole mass -- bulge mass correlation (see below); the 
match to the colour normalisation of the blue and red sequences is 
considerably improved (as a result of the higher yield adopted);
and the model reproduces the observed evolution of the luminosity function and
mass function at a similar level of success to the Bower et al.\ 2006 model. 
Further details are given in Bower et al.\ 2008.

\section{Why it works: Gas Mass Fractions}

\begin{figure}
  \includegraphics[height=8.4cm, angle=270]{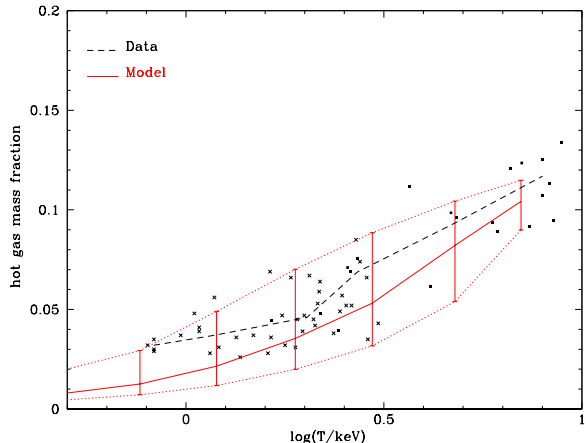}
\caption{The variation in hot gas mass fraction as a function of spectroscopic
temperature for the Bower et al.\ 2008 model. The red line shows the 
median X-ray emitting gas mass fraction of the model haloes.
The scatter in model is show by the error bars and dotted lines, which 
show the 10 to 90 percentile range in each temperature bin.
Measurements of the hot gas mass fraction are shown as solid black points, 
while the dashed black line shows the median fit to the observational data.}
\label{fig:fgas}
\end{figure}

As we have outlined in the previous section, the model's successful match to the
observed X-ray luminosities of groups and clusters is achieved by ``ejecting''
a large fraction of the hot baryons from the X-ray emitting regions of the lower
temperature haloes. 
In Fig.~\ref{fig:fgas} we focus on the hot gas mass fraction of the 
higher mass systems, where we can compare the model gas fractions 
with observational data. In order to do this, we
show the ratio of the hot gas mass to the total mass of the system as a function
of the system's X-ray temperature. The data (solid black points) are taken 
from profile measurements for clusters from Vikhlinin et al.\ (2006) and 
Pratt et al.\ (2006), and presented at an overdensity of 2500. These are 
supplemented by observations of 
galaxy groups taken from Sun et al.\ (2008), again within an overdensity of 2500.
We compare these measurements with the prediction of the model at the 
same overdensity. This region accounts for more than 70\% of the X-ray 
luminosity of the system and thus the gas fraction is closely tied to
the success of the model in accounting for the observed L--T relation. 

The median dependence in the model is shown
as a solid red line with error bars showing the scatter.
The model agrees reasonably well, and the trend of rapidly rising gas fraction 
around $T_{\rm spec}\sim3\keV$ is seen in both the data an the model. At lower
temperatures, the data tend to suggest somewhat higher gas fractions than
predicted by the model. However, the data shown here is probably biased to the 
most X-ray luminous galaxy groups. 
The plot also emphasises the large scatter in the hot gas mass fractions
of haloes at a given mass. The scatter in the model appears quite comparable to
that seen in the data. It will be intriguing to see how this comparison holds
up as the sample sizes increase and the sample selection becomes more 
representative.

\section{Discussion and Conclusions}

At the start of this talk, I set the challenge to build a model that was 
simultaneously able to account for the observed properties of 
galaxies such as the luminosity function and the X-ray properties of groups and clusters.
In this way we are seeking a model that accounts for both the number
and distribution of stars in the universe and the thermodynamic properties
of the material that is left over from the galaxy formation process. 
The importance of this second aspect is emphasised by looking at the 
mass fraction that it contains: in galaxy clusters more than 90\%
of the baryons are left over as a ``waste-product''. 

To rise to the challenge of modeling the ICM, we have made a relatively 
simple modification 
to the highly successful galaxy formation model presented in Bower et al.\ 2006. 
We have taken 
the radio-mode feedback process a step further, allowing the ``radio
mode'' of AGN feedback to eject gas from the X-ray emitting regions of hydrostatic 
haloes.
This modification of the model is successful in reproducing the 
observed correlations between X-ray luminosity and system temperature. In 
particular, as well as reproducing the median slope of the relation, the 
model reproduces the large scatter in the observed luminosities of lower
temperature systems. 
The steepening of the L--T relation results from gas
being ejected from the X-ray emitting regions of lower temperature groups,
and the diverse formation histories of these systems drives the 
large scatter in X-ray properties. The gas mass fractions of the model systems
agree well with the observed trends. 

A number of issues require closer examination, however.
 Firstly, we need to consider the
distribution of excess gas entropies and to be able to propagate this
through the merger hierarchy.  For example, a 
low mass substructure that has experienced little non-gravitational heating
tends to drop to the centre of a higher mass halo into which it is accreted 
with little increase in entropy. 
In this way it is quite simple to produce a small mass of low entropy 
(rapidly cooling) material
embedded in a halo of high entropy (long cooling time) gas. 
Secondly, the model needs to allow for the possibility that AGN activity might eject
some gas from haloes even if they are not in the hydrostatic regime. 
Both of these effects will tend to amplify the energy input
at earlier epochs making, reducing the energetic demands of the model
(although I should stress that the current model is already entirely consistent with
observed black hole masses and likely accretion efficiencies). 

In summary, the current model demonstrates that we are well
on the way to understanding physical process that set the combined properties 
of the galaxy population and the thermodynamic history of the intra-cluster
medium.  Given the high quality of current X-ray data, the flip-side of galaxy
formation provides important information that modeler's should cease to ignore!

%%%%%%%%%%%%%%%%%%%%%%%%%%%%%%%%%%%%%%%%%%%%%%%%
%% BACKMATTER
%%%%%%%%%%%%%%%%%%%%%%%%%%%%%%%%%%%%%%%%%%%%%%%%

\begin{theacknowledgments}
I thank the Galform team, particularly Andrew Benson and Ian McCarthy, without who this
development would not have been possible; and the organisers of the meeting for giving me 
chance to highlight this important aspect of galaxy formation models.
\end{theacknowledgments}

%%%%%%%%%%%%%%%%%%%%%%%%%%%%%%%%%%%%%%%%%%%
%% The following lines show an example how to produce a bibliography
%% without the help of the BibTeX program. This could be used instead
%% of the above.
%%%%%%%%%%%%%%%%%%%%%%%%%%%%%%%%%%%%%%%%%%%

\end{document}